\documentstyle[aps,prl,epsf]{revtex}

\begin{document}

\twocolumn[\hsize\textwidth\columnwidth\hsize\csname
@twocolumnfalse\endcsname

\draft
\title{``Superconductor"-insulator transitions in a Hubbard chain with
nearest-neighbor and bond-charge interactions}
\author{A. A. Aligia and E. R. Gagliano}
\address{Comisi\'on Nacional de Energ{\'\i}a At\'omica,\\
Centro At\'omico Bariloche and Instituto Balseiro,\\
8400 S.C. de Bariloche, Argentina.}
\date{Received \today }
\maketitle

\begin{abstract}
We consider a half-filled generalized Hubbard chain with electron-hole
symmetric correlated hopping and on-site and nearest-neighbor repulsions $U$
and $V$ respectively. In addition to the insulating charge- and spin-density
wave phases for large $V$ and $U$ respectively, we identify a phase with
dominant superconducting correlations at large distances for small $U$ and $V $. Using two Berry phases (one associated to the charge and the other to
the spin degrees of freedom) as discrete order parameters, we construct a
phase diagram for the three thermodynamic phases.
\end{abstract}


\vskip2pc]

\narrowtext

After the discovery of high-$T_{c}$ systems, there has been an increase in
the interest on purely electronic models exhibiting superconductivity,
particularly those with correlated hopping \cite
{yan,sud,lin,ar1,kha,san,jap,afq,c,mi}. A hopping dependent on the
occupation of the two sites involved is to be expected in nature, and
together with the on-site $(U)$ and nearest-neighbor $(V)$ repulsions, have
been discussed already by Hubbard in 1963 \cite{hub,stra}. The resulting
extended model (and a related one\cite{sev1}) has been studied by several
groups\cite{stra,sev,ov,ali,mo}, with particular emphasis in phase
transitions like the Mott metal-insulator one. At half-filling and when the
hopping term which changes the double occupancy $t_{AB}=0$, Strack and
Vollhardt obtained two regions of parameters for which the exact ground
state in arbitrary dimension is a charge-density wave (CDW) or a
spin-density wave (SDW) with maximum order parameter\cite{stra}. These
regions were later extended\cite{ov,ali} and a phase diagram constructed\cite
{ali}. In one dimension (1D) the result is represented by the dashed line in
Fig. 1 (b),(c). However, the nature of the third thermodynamic phase present
in the diagram has not been clearly elucidated, no exact result exist for $%
t_{AB}\neq 0\neq V$, and numerical studies of finite systems are not able to
identify sharply the phase boundaries\cite{mo,hir}. This difficulty is
mainly due to the fact that all correlation functions of finite systems vary
smoothly at the transition.

On the other hand, the interest on applications of geometrical Berry phases
has increased in recent years\cite{san,zak,res2,res,koi,gai,ort}. In
particular, Resta and Sorella have discovered that a Berry phase $\gamma _c$
(defined below) can be used as an order parameter for the ferroelectric
transition in a diatomic chain\cite{res}. In 1D problems with inversion
symmetry (like ours) $\gamma _c$ can only attain the values $0$ or $\pi $
(modulo $2\pi $)\cite{zak}. This quantization of $\gamma _c$ in only {\em two%
} possible values does not allow to distinguish between {\em three}
thermodynamic phases. Nevertheless,  we define another Berry
phase $\gamma _s$, with similar features as $\gamma _c$, and show that there
is a change in at least one of the topological numbers $\gamma _c/\pi
,\;\gamma _s/\pi $ at the transition between any two of the thermodynamic
phases. Therefore, each thermodynamic phase is univocally determined by a
two-component Berry phase vector ${\bf \gamma }=(\gamma _c,\gamma _s)$. This
fact allows us to define precisely all phase boundaries for any chain,
small to obtain its ground state by the Lanczos method
, and the phase diagram is obtained by finite-size scaling. In 1D the
nature of a thermodynamic phase is determined by the correlation functions
(charge-charge, spin-spin or pair-pair) with the lowest decay at large
distances. Calculation of the correlation exponents and other arguments
allow to identify the thermodynamic phase other than the CDW or SDW ones, as
``superconducting'' (S), in the sense that the system behaves as a
Tomonaga-Luttinger liquid in which the superconducting pair-pair
correlations are the greatest at large distances.

The Hamiltonian for a ring of $L$ sites with electron creation operators
satisfying arbitrary boundary conditions for both spins ${\bar{c}}%
_{j+L\sigma }^{\dagger }=e^{i\Phi _{\sigma }}{\bar{c}}_{j\sigma }^{\dagger }$
can be written in terms of the operators $c_{j\sigma }^{\dagger }=e^{-ij\Phi
_{\sigma }/L}{\bar{c}}_{j\sigma }^{\dagger }$ satisfying ${c}_{j+L\sigma
}^{\dagger }={c}_{j\sigma }^{\dagger }$ as: 
\begin{eqnarray}
H(\Phi _{\uparrow },\Phi _{\downarrow }) &=&\sum_{j\sigma }\{-e^{i\frac{\Phi
_{\sigma }}{L}}c_{j+1\sigma }^{\dagger }c_{j\sigma }(t_{AA}(1-n_{i,{\bar{%
\sigma}}})(1-n_{i+1,{\bar{\sigma}}})  \nonumber \\
&&+t_{AB}[n_{i,{\bar{\sigma}}}(1-n_{i+1,{\bar{\sigma}}})+n_{i+1,{\bar{\sigma}%
}}(1-n_{i,{\bar{\sigma}}})]  \nonumber \\
&&+t_{BB}n_{i,{\bar{\sigma}}}n_{i+1,{\bar{\sigma}}})+h.c.\}  \nonumber \\
&&+U\sum_{j}n_{j\uparrow }n_{j\downarrow }+V\sum_{j\sigma \sigma ^{\prime
}}n_{j+1\sigma }n_{j\sigma ^{\prime }}
\end{eqnarray}

In 2D, this Hamiltonian has been derived as an effective one-band model for
cuprate superconductors\cite{one}. We restrict the parameters to $%
t_{AA}=t_{BB}=1$, $0<t_{AB}\leq 1$, interpolating between the usual extended
Hubbard model $t_{AB}=1$\cite{hir}, and the case ${\it t}_{AB}=0$, for which
some exact results are known\cite{ar1,afq,stra,ov,ali}, particularly the
boundaries of the CDW and SDW phases\cite{ali} and the exact ground state
for $V=0$ \cite{ar1,afq}. Unfortunately this state is rather pathological,
highly degenerate, and at half-filling it is insulating even for small $U$,
in spite of the absence of a charge gap for $U<4$, for geometrical reasons%
\cite{afq,ali}. Thus, it is necessary to include the effect of $t_{AB}$ to
draw general conclusions on the nature of the Mott or
superconductor-insulator transition.

Due to the periodic boundary conditions used, the many-body eigenstates of $%
H $ for any fluxes $\Phi _\sigma $, can be classified according to their
total wave vector $K$ with $KL/2\pi $ integer (instead, the wave vectors $%
\bar{K}$ in the representation of the ${\bar{c}}_j^{\dagger }$ are shifted
and cover all real values as the $\Phi _\sigma $ are varied\cite{afq,res,ort}%
). If an eigenstate $|e_K(\Phi _{\uparrow },\Phi _{\downarrow })\rangle $ of 
$H(\Phi _{\uparrow },\Phi _{\downarrow })$ with a given $K$ does not have an
accidental degeneracy with another eigenstate with the same $K$ in the
interval $0\leq \Phi _{\uparrow }=\Phi _{\downarrow }=\Phi \leq 2\pi $, \cite
{note} then this state can be followed adiabatically as $\Phi $ is varied,
and a many-body generalization of Zak's Berry phase\cite{zak} can be defined
for it \cite{ort}. For the explicit calculation, it is more convenient to
use the numerically gauge invariant expression \cite{res2,ort}, discretizing
the interval $0<\Phi <2\pi $ into $N$ points $\Phi _i=2\pi i/N$: 
\begin{eqnarray}
\gamma _c(e_K) &=&-\lim_{N\rightarrow \infty }\{\text{Im}[\ln (\Pi
_{i=0}^{N-2}\langle e_K(\Phi _i,\Phi _i)|e_K(\Phi _{i+1},\Phi _{i+1})\rangle
\nonumber \\
&&\langle e_K(\Phi _{N-1},\Phi _{N-1})|e_K^c(2\pi )\rangle )]\},
\end{eqnarray}
\noindent where $|e_K^c(2\pi )\rangle $ represents $|e_K(2\pi ,2\pi )\rangle 
$ obtained directly from $|e_K(0,0)\rangle $\cite{note2}: 
\begin{equation}
|e_K^c(2\pi )\rangle =\exp [i{\frac{{2\pi }}L}\sum_{j\sigma }j{c}_{j\sigma
}^{\dagger }{c}_{j\sigma }]|e_K(0,0)\rangle .
\end{equation}
\noindent In practice $N=10$ points are more than enough to define if $%
\gamma _c(e_K)$ is $0$ or $\pm \pi $ with six digits accuracy. In few cases
in which this was not the case, we increased $N$ until the same level of
accuracy was obtained. Note that in contrast to previous cases\cite{res,ort}%
, $|e_K(\Phi ,\Phi )\rangle $ is not necessarily the ground state. In
particular, for most of the parameters inside the S phase, there is a
crossing of the energy levels between $K=0$ and $K=\pi $. As a consequence,
the ground-state energy $E_g(\Phi ,\Phi )$ has local minima around $\Phi =0$
and $\Phi =\pi $ (see Fig. 2) and its form is suggestive of anomalous flux
quantization (AFQ), $i.e.$ $E_g(\pi )=E_g(0)$ in the thermodynamic limit, as
expected if superconducting correlations dominate at large distances\cite
{sud,afq}.

In analogy to the fact that from the ground-state energy $E_g(\Phi
_{\uparrow },\Phi _{\downarrow })$ one can calculate a charge stiffness
(Drude weight) and a spin stiffness as $D_c=(L/2)\partial ^2E_g(\Phi ,\Phi
)/\partial \Phi ^2$, $D_s=(L/2)\partial ^2E_g(\Phi ,-\Phi )/\partial \Phi ^2$%
, for $\Phi=0$, respectively\cite{afq}, we define the spin Berry phase as
that captured varying $\Phi _{\uparrow }$ from $0$ to $2\pi $, keeping $\Phi
_{\downarrow }=-\Phi _{\uparrow }$ in the cycle:

\begin{eqnarray}
\gamma _{s}(e_{K}) &=&-\lim_{N\rightarrow \infty }\{\text{Im}[\ln   \nonumber
\\
&&(\Pi _{i=0}^{N-2}\langle e_{K}(\Phi _{i},-\Phi _{i})|e_{K}(\Phi
_{i+1},-\Phi _{i+1})\rangle   \nonumber \\
&&\times \langle e_{K}(\Phi _{N-1},-\Phi _{N-1})|e_{K}^{s}(2\pi )\rangle
)]\},
\end{eqnarray}

\begin{equation}
|e_K^s(2\pi )\rangle =\exp [i{\frac{{2\pi }}L}\sum_{j\sigma }j\sigma {c}%
_{j\sigma }^{\dagger }{c}_{j\sigma }]|e_K(0,0)\rangle ,
\end{equation}
\noindent where $\sigma =1$(-1) for spin up(down). While in general $\gamma
_c$ is a measure of the polarization in the system \cite{zak,res2,res,ort}
(a change of $\gamma _c$ in $\pi $ corresponds to the change of polarization
when one unit charge is displaced one lattice parameter), $\gamma _s$ is a
measure of the difference between the polarization of electrons with spin up
and down.

At this point it is interesting to note the effect of the unitary
transformation, 
\begin{equation}
{c}_{j\uparrow }^{^{\prime }}={c}_{j\uparrow },\;\;{c}_{j\downarrow
}^{^{\prime }}=(-1)^j{c}_{j\downarrow }^{\dagger },
\end{equation}

\noindent on $H$. In the resulting Hamiltonian $H^{\prime }$, $\Phi
_{\downarrow }$ and $U$ change sign, while the term in $V$ transforms to an
Ising interaction $\hat{V}^{\prime }=V(n_{j\uparrow }-n_{j\downarrow
})(n_{j+1\uparrow }-n_{j+1\downarrow })$. Assuming in the following $L$
even, and calling $|e_{K}^{^{\prime }}\rangle $ the state obtained from $%
|e_{K}\rangle $ through the unitary transformation one has: 
\begin{equation}
\gamma _{s}(e_{K}^{^{\prime }})=\pi +\gamma _{c}(e_{K})
\end{equation}
\noindent The term $\pi $ is due to the fact that the transformation
interchanges the vacuum with the fully polarized ferromagnetic state $%
|F\rangle ,$ for which $\gamma _{c}(F)=\gamma _{s}(F)=\pi $. This can be
easily seen from Eqs. (2)-(5), since all scalar products inside the
logarithm are 1, except the last one, which is -1.

We define the Berry phases of a thermodynamic phase, as the Berry phases of
the corresponding ground state. For those cases in which two energy levels
with different quantum numbers as functions of $\Phi _{\uparrow },\Phi
_{\downarrow }$ cross in the ground state, we have chosen the state of
lowest minimum of the energy $E(\Phi _{\uparrow },\Phi _{\downarrow })$. In
all cases we have found that this minimum corresponds to $K=\Phi _{\uparrow
}=\Phi _{\downarrow }=\pi $ for $L$ multiple of four, and to $K=\Phi
_{\uparrow }=\Phi _{\downarrow }=0$ if $L/4$ is a half integer. The chosen
state has also the lowest energy averaged over the fluxes. The values of the
Berry phases at the CDW and SDW thermodynamic phases are easy to determine
using continuity arguments ($\gamma _s,\gamma _c$ cannot jump unless there
is a phase transition). In the limit of very large $U$, for $\Phi _{\uparrow
}=\Phi _{\downarrow }=\Phi $, $H$ is equivalent to a Heisenberg Hamiltonian
with exchange constant $J=4t_{AB}^2/(U-V)$ independent of $\Phi $. Then, all
factors but the last in Eq. (2) are 1, the exponential in Eq. (3) introduces
a factor -1, and $\gamma _c($SDW$)=\pi $. A Hartree-Fock wave function leads
to the same result, in agreement with previous findings for a diatomic system%
\cite{ort}. When $\Phi _{\uparrow }=-\Phi _{\downarrow }$, the exchange
terms in the large $U$ limit depend on $\Phi _{\uparrow }$. However we do
not expect a transition if the Ising term (of the form of $\hat{V}^{\prime }$%
) is increased (artificially). This favors a Neel configuration and Eqs. (4)
and (5) lead to $\gamma _s($SDW$)=\pi $. Defining a vector ${\bf \gamma }%
=(\gamma _c,\gamma _s)$, we can write ${\bf \gamma }($SDW$)=(\pi ,\pi )$.
Similarly in the large $V$ limit, the ground state is the CDW with maximum
order parameter and from Eqs. (2)-(5), it easily follows that ${\bf \gamma }%
( $CDW$)=(0,0)$. This result is consistent with Eq. (7) and with the changes
in polarization for both spins when in the Neel SDW state, all electrons
with spin up are displaced one lattice parameter to form a CDW state. These
simple arguments cannot be extended to the S phase near $U,V\sim 0$. However
for $U=V=0$, and $0\neq t_{AB}\neq 1$, the ground state is non-degenerate
for any fluxes $\Phi _{\uparrow },\Phi _{\downarrow }$, and therefore, it
should be mapped onto itself by the transformation Eq. (6). Then, Eq. (7)
implies that $\gamma _s($S$)=\pi +\gamma _c($S$)$ (modulo $\pi $). Thus,
before actually computing the Berry phases one knows that the ${\bf \gamma }$
are different for the three phases. We obtain ${\bf \gamma }($S$)=(0,\pi )$
numerically. In the non-interacting case $U=V=0$, $t_{AB}=1$, ${\bf \gamma }$
is ill defined due to a degeneracy of the ground state for some values of $%
\Phi _{\uparrow },\Phi _{\downarrow }$ (corresponding to a local maximum in
the ground-state energy as a function of flux) and the same value of $K$.
This is consistent with previous numerical calculations, which found that $%
\gamma _c$ in a metallic normal phase took arbitrary values and did not
converge with the number of splittings $N$\cite{ort}.

\begin{figure}
\narrowtext
\epsfxsize=3.3truein
\vbox{\hskip 0.05truein \epsffile{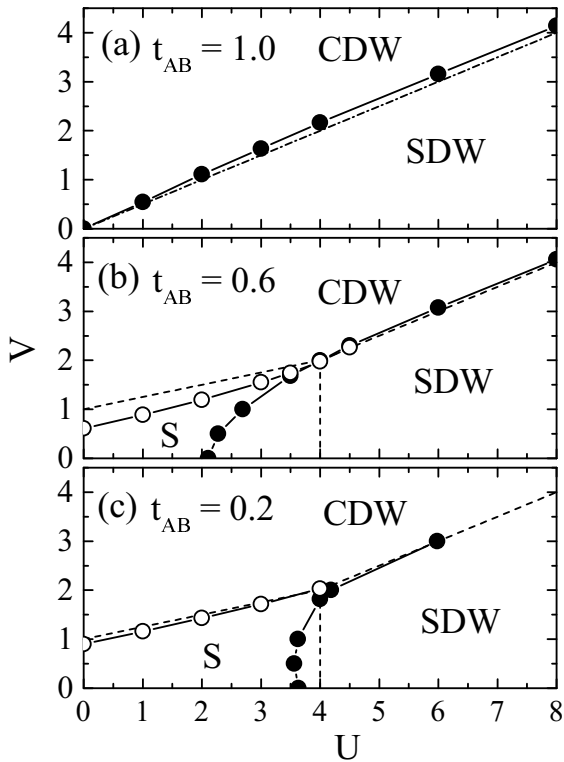}}
\medskip
\caption{Phase diagram in the $U,V$ plane for different
values of $t_{AB}$. Solid (empty) circles denote the points at which $\gamma
_c$ ($\gamma _s$) jumps in $\pi $. Dashed line corresponds to the exact
result for $t_{AB}\rightarrow 0$ {\protect \cite{ali}}. Dashed-dotted line in (a)
corresponds to $V=U/2$.}
\end{figure}

The phase diagram for different values of $t_{AB}$ is shown in Fig. 1.
Because of the unitary transformation Eq. (6), the phase diagram is also
that of $H^{\prime }$ interchanging SDW and CDW phases. The results were
obtained calculating the boundaries for all chains of even length $L\leq 10$
and extrapolating the results using the procedure followed in Ref.\cite{fou}%
. For $t_{AB}=1$, the S phase is absent and the CDW-SDW boundary is very
similar to that found previously from the structure factors obtained with
Monte Carlo simulations\cite{hir}. However, this method has an error $\sim
0.05$ in the critical value of $V$. There is in addition a 
systematic error due to the discretization of imaginary times which is
difficult to estimate. In our case,
from the difference in the extrapolated values using up to 8 sites, or up to
10 sites, we estimate the error to be one order of magnitude smaller. As
long as $t_{AB}<1$, a superconducting phase appears, near the CDW-SDW
boundary for $t_{AB}=1$, if $U<4$, particularly for small $U$ and $V$. This
can be understood in part by the fact that the energy of CDW and SDW states
increase when $t_{AB}$ decreases. This fact is particularly clear for large $%
V$ or large $U$ for which the ground-state energy per site is $%
E_{CDW}=U/2-2t_{AB}^{2}/(3V-U)$ and $E_{SDW}=V-4t_{AB}^{2}\ln 2/(U-V)$
respectively, and agrees roughly with the result of a generalized
Hartree-Fock decoupling of the two- and three-body terms present in $H$\cite
{mo}. The exact results for $t_{AB}=0$ \cite{ar1,afq,stra,ov,ali}, show that
for small $t_{AB},U$ and $V$, it is energetically convenient to destroy the
CDW or SDW phase to take advantage of the kinetic energy terms in $t_{AA}$
and $t_{BB}$. However, the fact that the S phase is more stable near the
line in which the CDW and SDW are degenerate is noticeable and unexpected in
these previous approaches. This is probably related to the fact that the
transition between the CDW and SDW phases for $t_{AB}=1$ is second order for 
$U<U_{c}$ and first-order for $U>U_{c}$, where the position of the
tricritical point $U_{c}\sim 3\pm 1$\cite{hir}. The S phase seems to develop
around the line of the second-order phase transition as long as $t_{AB}<1$.
Unfortunately, our method can not determine the order of the transition and
a more precise value of $U_{c}$ can not be given.

As $t_{AB}$ decreases further, the region occupied by the S phase increases
and tends to the exact result for $t_{AB}\rightarrow 0$\cite{ali}. For small 
$t_{AB}$ the finite-size effects become more important for the S-SDW phase
boundary. For $V=0$ and $t_{AB}=0.1$, the extrapolation using chains up to $%
10$ sites gives $U_{S-SDW}=3.901$, while extrapolation including $L=12$
gives a value smaller by 0.016.

\begin{figure}
\narrowtext
\epsfxsize=3.3truein
\vbox{\hskip 0.05truein \epsffile{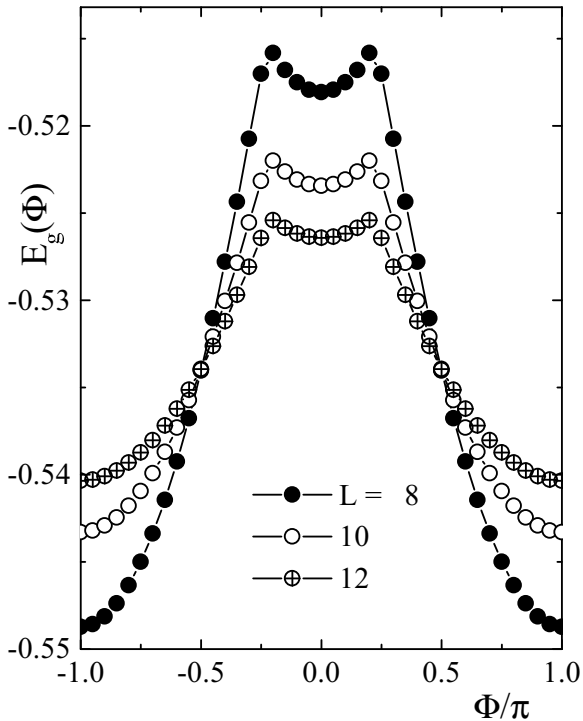}}
\medskip
\caption{Ground-state energy as a function of the flux $%
\Phi _{\uparrow }=\Phi _{\downarrow }=\Phi $ for $U=1.5$, $V=0$, $t_{AB}=0.5$
and several values of $L$.}
\end{figure}

Previous numerical studies for finite $t_{AB}$, well inside the S phase have
shown the presence of mobile carriers\cite{mo}, however the attempts to
determine if the S phase is superconducting or not gave contradictory results%
\cite{c}. On one hand, it has been established that for $t_{AB}<1$, and
small $U$ and $V$, the system behaves as a Tomonaga-Luttinger liquid with
dominant superconducting correlations at large distances. On the other hand,
no signs of AFQ were found for $U=V=0$, and a generalized Hartree-Fock
Bardeen-Cooper-Schrieffer (BCS) decoupling of the two- and three-body terms,
shows that an ordinary singlet BCS superconducting solution is unstable.
However, we find that this decoupling, for small $U,V$, leads naturally to a 
$p$-wave triplet-pairing superconducting ground state. A similar BCS
solution exists in 2D, which should be valid for small $U,V$ and $1-t_{AB}$.
In addition, since the dependence of the ground-state energy with the flux
is calculated simultaneously with the Berry phase $\gamma _c$, we found that
in most of the S phase there are signs of AFQ. An example is given in Fig.
2. The fact that no signals of AFQ are found for $U=V=0$ is probably a
finite-size effect. Thus, all calculations are consistent with the
superconducting nature of the S phase. Furthermore, for $V=0$, the
correlation exponent $K_\rho >1$ for $U<U_{S-SDW}$, while for $U>U_{S-SDW}$
a gap opens in the charge sector \cite{c}. Calculations of $K_\rho $ for $%
t_{AB}=0.6$ in rings of 8, 10 and 12 sites gave $U_{S-SDW}=2.05\pm 0.05$, in
excellent agreement with the result 2.11 obtained with the Berry phases (see
Fig. 1 (b)). The triplet nature of the S phase is supported by our numerical
calculations of pair-pair correlation functions for $L=10$ and the fact that
for a singlet S phase one expects $\gamma=(0,0)$.

In summary, we have shown that in one dimension at half filling, the
extended Hubbard model with correlated hopping $H$ (Eq. (1)), and the model $%
H^{\prime }$ obtained from $H$ through the unitary transformation Eq. (6),
present a phase ``S'' with dominating superconducting correlations at large
distances for $0<t_{AB}<t_{AA}=t_{BB}$ and small $U$ and $V$. This
thermodynamic phase, as well as the insulating CDW and SDW phases , which
are the ground state for large $V$ and large $U$ respectively, are
characterized by different quantized values of ${\bf \gamma }=(\gamma
_c,\gamma _s)$, where $\gamma _c,\gamma _s$ are Berry phases associated with
charge and spin degrees of freedom respectively: ${\bf \gamma }($S$)=(0,\pi
),~{\bf \gamma }($CDW$)=(0,0),~{\bf \gamma }($SDW$)=(\pi ,\pi )$. The finite
jumps in ${\bf \gamma }$ at the phase boundaries allowed us to define them
sharply for each finite chain, and to obtain an accurate phase diagram
through finite-size scaling. The use of topological quantum numbers to
characterize the different thermodynamic phases should be useful in general
for non-trivial highly correlated models, for which correlation functions in
finite systems do not show sharp transitions and exact solutions in the
thermodynamic limit are very rare.

One of us ( A.A.A. ) wants to thank Sandro Sorella for discussions held at
the Miniworkshop on Strong Electron Correlations at ICTP, Trieste, Italy (
July 1996), and bringing Ref.\cite{res2} to our attention. Partial support
from Agencia Nacional de Promoci\'{o}n Cient\'{\i}fica y Tecnol\'{o}gica
under grant PMT-PICT0005 is gratefully acknowledged. One of us (E.R.G.) is
supported by CONICET, Argentina. A.A.A. is partially supported by CONICET.




\begin{references}
\bibitem{note}  {For some parameters and values of $K$, there are accidental
degeneracies at given values of $\Phi _{\downarrow }=\Phi
_{\uparrow }$ ($\Phi _{\downarrow }=-\Phi _{\uparrow }$%
) and precisely at these points $\gamma _c$ ($\gamma _s$) jumps in $\pi $.}
\bibitem{note2}  {In terms of the ${\bar{c}}_{j\sigma }^{\dagger }$
operators, $|e_K^c(2\pi )\rangle $ is identical to $|e_K^{}(0,0)\rangle $.
Also for $\Phi =0$, ${\bar{c}}_{j\sigma }^{\dagger }={c}_{j\sigma }^{\dagger
}$. Then, writing $|e_K^c(2\pi )\rangle $ in terms of the ${c}_{j\sigma }$
operators for $\Phi =2\pi $ leads to Eq. (3).}
\end{references}
\end{document}